# Stability-Drift Early Warning for Cyber-Physical Systems Under Degradation Attacks

Daniyal Ganiuly, Nurzhau Bolatbek, Assel Smaiyl
Astana IT University, Astana, Kazakhstan

**Abstract:** Cyber-physical systems (CPS) such as unmanned aerial vehicles are vulnerable to slow degradation that develops without causing immediate or obvious failures. Small sensor biases or timing irregularities can accumulate over time, gradually reducing stability while standard monitoring mechanisms continue to report normal operation. Detecting this early phase of degradation remains a challenge, as most existing approaches focus on abrupt faults or visible trajectory deviations. This paper introduces an early warning method based on stability drift, which measures the divergence between predicted and observed state transitions over short horizons. By tracking the gradual growth of this divergence, the proposed approach identifies emerging instability before it becomes visible in the flight trajectory or estimator residuals. The method operates externally to the flight stack and relies only on standard telemetry, making it suitable for deployment without modifying autopilot firmware. The approach was evaluated on a PX4 x500 platform in a software in the loop environment under two realistic degradation scenarios, gradual IMU bias drift and timing irregularities in the control loop. In both cases, the stability drift metric provided a consistent early warning signal several seconds before visible instability appeared, while remaining stable during nominal and aggressive but non degraded flight. The results demonstrate that stability drift can serve as a practical indicator of early degradation in UAV control systems. By providing advance notice during a pre instability phase, the proposed method complements existing safety mechanisms and offers additional time for mitigation or safe mode transitions under slow and subtle attacks.

**INTRODUCTION**
Cyber-physical systems rely on tight interaction between sensors, computation, and control logic. Unmanned aerial vehicles are a clear example because they must constantly adjust their motion based on real-time sensor data. Even small disturbances can influence the stability of the entire control loop. These disturbances are not always sudden or easy to notice. Some grow slowly during flight and cause a gradual loss of stability. This slow form of degradation is especially concerning because it does not look harmful at the beginning. The drone follows its planned trajectory, the estimator produces reasonable values, and none of the usual monitoring signals change [1].
Most security mechanisms for drones focus on faults or attacks that create visible deviations. They rely on large estimation errors, strong jumps in sensor values, or clear oscillations in the controller. These methods work well when the drone experiences a sharp disruption. They do not work as well when the disturbance grows slowly [2]. For example, a small increase in accelerometer bias stays hidden for a long time. At first it has almost no effect on the path or on the estimated state. Timing jitter behaves in a similar way. A short delay in the control loop does not break stability immediately [3]. The problem appears only after many cycles, when the accumulated delay begins to influence the motion of the drone. In our experiments with the PX4 x500 model in a software-in-the-loop environment, we observed that slow attacks create very early signs of instability. These signs do not appear in the flight path or in the residuals produced by the estimator. They appear only when we compare the predicted state of the drone with the state measured by its sensors. The difference between these two states grows gradually and becomes noticeable long before the drone shows any visible deviation [4]. We refer to this gradual change as stability drift. Stability drift is a reliable indicator that the system is moving toward a condition where the controller will no longer behave as expected, even though all basic monitoring tools still consider the system to be normal.

Previous research on drone security has examined sensor spoofing, timing faults, parameter changes, and unexpected controller behavior. These studies have shown that drones are sensitive to disturbances that affect their control loops, but the detection methods mostly react after the system has already begun to degrade. Other work has explored fuzzing of autopilot configurations or runtime defenses that help the drone recover from failure. These approaches reveal weaknesses or enable recovery, but they do not provide a mechanism to recognize the beginning of degradation. As a result, the system continues to operate without awareness of its growing instability.

Our goal in this work is to close this gap by developing a method that identifies degradation in its earliest stage. We introduce a stability drift early warning system that monitors the divergence between predicted and actual state transitions. The system uses a short sliding window of prediction errors and raises an alert when it detects a steady increase. It does not wait for visible oscillation or a clear change in the trajectory. It reacts to the first signs of abnormal dynamics. The design works as an external module and does not require changes to the PX4 firmware.

We evaluate the method on two slow attacks that reflect realistic threats to drone control. The first is a gradual drift in accelerometer and gyroscope bias that increases during flight. The second is timing jitter created by irregular delays in the control loop. Both attacks were tested on the PX4 x500 platform in the Gazebo simulator while the drone followed a square path at a constant altitude and speed. In both scenarios the stability drift metric reacted significantly earlier than a residual-based detector. For the bias drift attack our method raised an alert after about twelve seconds, whereas the baseline detector reacted only after a clear tracking error appeared near the forty-second mark. The timing jitter experiment showed a similar pattern, with early detection at roughly eight seconds, while the baseline remained quiet until the late stage of degradation.

## RELATED WORK

Research on the security of cyber-physical systems has produced a wide range of methods for identifying abnormal behavior in sensing and control loops. Work on UAV security has focused on three main directions: sensor spoofing detection, analysis of timing faults and scheduling delays, and runtime monitoring of controller behavior. Each direction provides useful insight into the behavior of UAVs under specific forms of disturbance, but none of these approaches identifies the early changes in system dynamics that appear during slow degradation attacks.

### A. Sensor spoofing detection

Several studies have examined how drones react to corrupted measurements from GPS, inertial sensors, barometers, or magnetometers. These methods typically rely on inconsistencies between sensor modalities or on noticeable deviations in the estimator's innovation signal [5]. When an attacker injects an abrupt offset into a sensor, the estimator quickly becomes inconsistent with the drone's predicted motion. Residual-based detectors respond well to these scenarios because the error grows rapidly [6]. However, these detectors assume that the attacker introduces a significant jump in the measurements. Slow bias drift does not produce a clear inconsistency for a long period of time, so these methods remain silent until the accumulated drift becomes large enough to influence the state estimate.

### B. Timing faults and control-loop delays

Recent work has shown that UAV controllers are sensitive to irregular timing. Studies on autopilot timing behavior, scheduler interactions, and event delays demonstrated that even small variations in update timing can affect the trajectory of a drone [7]. Several papers propose tools that identify these timing issues by stressing the scheduling layer or by measuring deadlines inside the flight stack. These methods rely on instrumenting the autopilot or observing violations of timing guarantees. They are effective for identifying faults that violate the expected update period [8]. They do not, however, track the gradual impact of timing jitter on the physical state of the drone. When the delay is small and spread over many cycles, the controller remains stable long enough that timing-based detectors do not generate warnings.

### C. Runtime monitoring and fault-tolerant control

Autopilot platforms such as PX4 and ArduPilot include built-in fail-safe mechanisms and fault-detection modules that monitor sensor validity and actuator limits. Research efforts have extended these components with runtime monitors that check for anomalous values or unexpected changes in the internal state of the autopilot [9]. These systems are designed to detect well-defined failure modes, such as a stalled sensor, a large jump in measurement values, or an actuator that saturates. They provide valuable protection against known faults but do not capture gradual physical effects that accumulate over time. The monitoring functions are intentionally conservative because they must avoid false alarms during normal flight, which makes them less sensitive to the early signs of degradation [10].

### D. Analysis of autopilot robustness and fuzzing of configuration parameters

Recent work has also explored the robustness of UAV flight stacks through fuzzing and stress testing. These approaches uncover software bugs, parameter interactions, and configuration issues that influence flight stability [11][12]. Tools in this space inject malformed inputs or vary system parameters to trigger edge cases inside the autopilot. Although this line of research reveals important weaknesses and helps evaluate resilience, it concentrates on software robustness rather than the physical evolution of the drone under subtle disturbances. It does not provide a mechanism for estimating the earliest point at which the physical system begins to diverge from its expected behavior [13].

Across all these directions, detection methods rely on either discrete anomalies in sensor or timing data or observable performance degradation. None of the existing approaches track the gradual change in physical stability that begins long before visible deviations appear. To our knowledge, no prior work has used short-horizon prediction errors to monitor the early drift between expected and actual dynamics in UAV control. The concept of stability drift has not been used as an early-warning signal, nor has it been evaluated on attacks that increase bias or introduce low-rate timing jitter [14]. This gap motivates the design of our stability-drift early warning system, which follows a different strategy from prior anomaly detection and runtime monitoring approaches.

## SYSTEM AND THREAT MODEL

### A. System Model

The system under study is a quadrotor running the PX4 flight stack in a software-in-the-loop environment. The drone model is based on the x500 frame and uses a standard multirotor configuration with four brushless motors, a three-axis accelerometer, a three-axis gyroscope, and a barometer for altitude estimation. PX4 calculates thrust and torque through motor mixing, and the dynamics of the vehicle are simulated in the Gazebo physics engine. This setup provides realistic inertial behavior, sensor noise, actuator latency, and aerodynamic effects.

PX4 uses an estimator that fuses IMU and other sensor data to obtain position, velocity, and attitude of the drone [15][16]. The estimator runs at a fixed rate of 250 Hz internally, while position and velocity estimates are made available over MAVLink at 50 Hz. The controller receives a new state estimate at each cycle, computes the required attitude and thrust, and sends actuator commands according to the requested flight mode. In our evaluation, the drone executes an autonomous mission using offboard waypoint tracking. This produces a consistent control-loop structure that is easy to monitor and repeat across experiments.

During normal operation, the estimator and controller assume that timing between cycles remains close to the expected period and that sensor noise stays within its nominal model. Any deviation from these assumptions alters the evolution of the state and eventually the stability of the control loop [17][18]. This sensitivity to small changes creates opportunities for slow degradation attacks.

As shown on Fig 1, telemetry data are transmitted from the UAV to an external monitoring process that observes system behavior and generates early-warning signals without interacting with flight control or estimator logic.

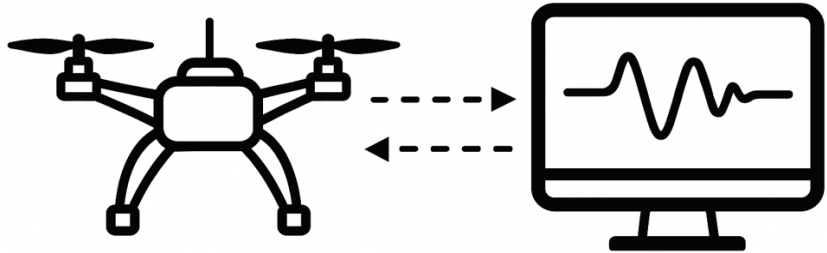

Fig 1. Conceptual overview of an externally monitored UAV

**B. Attacker Model**

We consider an adversary whose goal is to cause a gradual loss of stability in the drone without creating sudden or easily recognizable anomalies. The adversary does not attempt to crash the drone or force it into an obviously unsafe state. Instead, the attacker aims to create physical behavior that diverges slowly from the predicted motion, which makes early detection difficult for standard monitoring systems.

The attacker has the ability to influence either sensor outputs or the timing of the control loop. These capabilities reflect real vulnerabilities that may arise in practice. For example, a compromised sensor driver may introduce a small bias that evolves over time, and a malicious software component may delay control updates at irregular intervals. The attacker does not modify the PX4 firmware directly and does not inject large jumps in sensor values. The intention is to remain undetected for as long as possible.

We focus on two specific attack vectors:
- IMU bias drift: The attacker introduces a slowly increasing bias into the accelerometer or gyroscope readings. The bias is small enough that each individual sample appears normal. The estimator incorporates the corrupted readings into its state, which gradually shifts the controller's understanding of the drone's motion. Because the drift grows at a constant and low rate, traditional detectors do not observe any immediate inconsistency.
- Timing jitter in the control loop: The attacker introduces irregular delays into the controller update path. Each delay is short and infrequent, so the control loop remains functional. However, repeated delays over time disturb the regular timing of state updates. This affects the stability of the controller because its calculations assume a fixed update rate. The effect appears only after many cycles and is difficult to distinguish from normal variability.

In both cases, the attacker operates under a constraint: the disturbance must remain small enough that the drone maintains stable flight during the early phase of the attack. This constraint reflects realistic conditions in which an attacker seeks to avoid raising alarms while still causing long-term degradation [19][20]. The defender does not assume access to low-level autopilot modifications and relies only on telemetry streams available through the standard MAVLink interface.

# STABILITY DRIFT FORMULATION

The stability drift metric is designed to identify the earliest signs of physical degradation in the control loop. The method relies on monitoring how the drone's actual motion diverges from the motion predicted by a simplified dynamic model. This divergence begins to increase before the drone exhibits visible instability, which makes it a suitable early indicator of slow-developing attacks.

## A. Short-Horizon State Prediction

At each control cycle, the current state estimate $x_t$ and control input $u_t$ are received through the telemetry interface. A prediction of the next state $\hat{x}_{t+1}$ is then produced using a simplified quadrotor model. The model incorporates the main translational and rotational dynamics of a multirotor and uses the sample period of the control loop to advance the state one step forward. The prediction is computed according to:

$$\hat{x}_{t+1} = f(x_t, u_t)$$

The function $f(x_t, u_t)$ follows a standard discrete-time multirotor model incorporating thrust-based acceleration, rotational dynamics, and gravity. Its full expression is omitted as it follows conventional PX4 formulations and is not required for the proposed method. Although this model is not a complete reproduction of the PX4 internal dynamics, it is sufficiently accurate to reveal small deviations caused by timing irregularities or sensor drift.

## B. Model-Prediction Divergence

The deviation between the predicted state and the measured state is quantified by the model-prediction divergence $D_t$:

$$D_t = \|x_{t+1} - \hat{x}_{t+1}\|$$

## C. Stability Drift Metric

To reduce the influence of noise and short-term disturbances, the divergence values are aggregated over a sliding window of length k. The stability drift metric, denoted as SD_t, is defined as:

$$SD_t = \frac{1}{k} \sum_{i=t-k}^{t} D_i$$

This formulation highlights long-term trends in the divergence signal. When the system remains stable, the aggregated value fluctuates within a narrow range. When degradation begins, the values within the window rise steadily, even though the drone may still follow its planned trajectory without visible deviation.

## D. Dynamic Thresholding

A baseline distribution of stability drift values is established during nominal flight. The threshold for detecting abnormal behavior is computed as:

$$T = \mu_{SD} + \alpha \sigma_{SD}$$

Here, $\mu_{SD}$ denotes the mean of the stability drift metric during stable operation, $\sigma_{SD}$ denotes the standard deviation, and \alpha is a sensitivity parameter. The value $\alpha = 2.5$ provides a balance between early detection and robustness to normal fluctuations.

### E. Early-Warning Decision

An early-warning event is triggered when the stability drift exceeds the threshold:

$$SD_t > T$$

This condition indicates that the system's physical behavior has begun to diverge from its expected dynamics. The event does not represent an imminent failure but serves as advance notice that the stability margin of the control loop is decreasing. The signal can then be used by higher-level components to adjust mission parameters or activate protective mechanisms.

The formulation operates externally to the PX4 flight software and depends only on telemetry data accessible through MAVLink. This allows the method to be integrated into existing UAV monitoring systems without changes to the autopilot.

## EARLY WARNING ARCHITECTURE

The early-warning architecture operates alongside the flight stack and observes the system through standard telemetry streams. All computations are performed externally to the PX4 autopilot, which allows the detector to monitor the system without modifying control logic or estimator internals. State estimates and control signals are obtained through MAVLink and processed in real time.

Fig 2 illustrates how the proposed early-warning system operates alongside the flight stack. The PX4 autopilot remains unchanged and publishes state estimates and control signals through standard MAVLink telemetry, which are processed externally by the monitoring module.

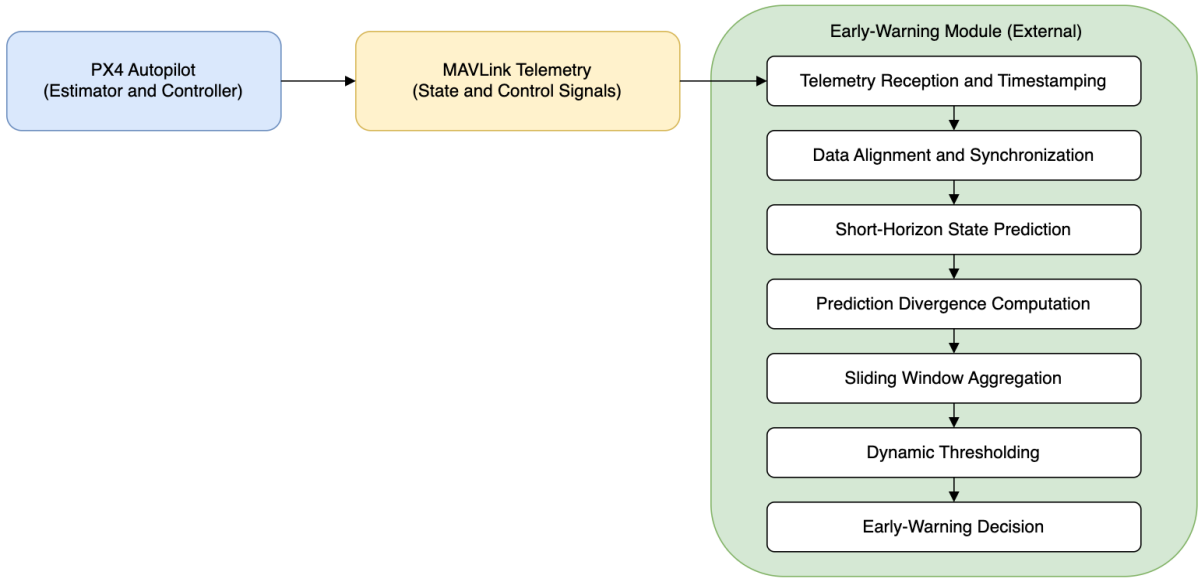

Fig 2. Runtime Architecture and Data Flow

### A. Runtime Data Flow

During flight, PX4 publishes state information and control outputs at a fixed rate. These messages include position, velocity, attitude, angular rates, and the corresponding control inputs applied by the controller. The monitoring process subscribes to these streams and assigns timestamps upon reception. Because telemetry messages may arrive with small timing variations, the incoming data are aligned to a common timeline before further processing. This alignment ensures that the state $x_t$, the control input $u_t$, and the subsequent state measurement $x_{t+1}$ correspond to the same control cycle. Once synchronized, the data are forwarded to the prediction module.

## B. Online Drift Monitoring

The prediction module computes a one-step state estimate using the simplified dynamic model described earlier. When the next telemetry update becomes available, the predicted and measured states are compared, and the resulting divergence value is added to a sliding window. This window is updated continuously and provides a smoothed view of how the system's behavior evolves over time.

Under nominal conditions, the aggregated drift metric remains within a narrow range. When degradation begins, small discrepancies accumulate and cause a gradual increase in the metric, even though the drone may still track its trajectory accurately. This behavior allows the system to detect emerging problems before they manifest as visible instability.

## C. Early-Warning Logic and System Interaction

The drift metric is evaluated against a dynamic threshold derived from nominal flight behavior. When the threshold is exceeded, an early-warning signal is generated. The signal reflects a sustained deviation between expected and observed dynamics rather than an abrupt fault.

The detection mechanism is intentionally separated from any control response. The early-warning signal can be consumed by a mission supervisor, a safety module, or a ground control station, depending on the deployment scenario. This separation prevents the detector from influencing control decisions directly and preserves compatibility with existing flight software.

## EXPERIMENTAL SETUP

The evaluation was conducted in a controlled simulation environment that allows realistic UAV behavior while enabling precise manipulation of sensing and timing conditions. A software-in-the-loop configuration was selected to ensure repeatability of experiments and full observability of system dynamics, which are required for analyzing gradual degradation effects.

## A. UAV Platform and Simulation Environment

All experiments were performed using PX4 Software-In-The-Loop (SITL), which executes the full PX4 flight stack, including state estimation, control logic, and actuator mixing. The simulated vehicle follows a standard quadrotor configuration consistent with commonly used PX4 reference models. This setup provides realistic closed-loop behavior while allowing controlled modification of sensor and timing characteristics.

The early-warning system was implemented as an external monitoring process running alongside the simulator on the same host machine. Communication between PX4 and the monitoring process was established through MAVLink over UDP. This configuration reflects a practical deployment scenario in which a companion computer or ground-based system observes the UAV during operation without requiring access to internal autopilot components.

## B. Telemetry and Monitoring Configuration

The monitoring process subscribed to MAVLink telemetry streams providing position, velocity, attitude, angular rates, and applied control inputs. These signals were sufficient to reconstruct the state vector required for stability-drift evaluation. Telemetry was processed continuously at the simulator's native update rate.

Incoming messages were timestamped and aligned to ensure that each state measurement, the corresponding control input, and the subsequent state update belonged to the same control cycle. This alignment step was necessary to maintain consistency in the presence of minor message timing variations. No internal PX4 variables, estimator states, or debugging interfaces were accessed. All monitoring relied exclusively on standard telemetry available on real UAV platforms.

## C. Degradation Attack Injection

Two classes of degradation attacks were considered: sensor bias drift and timing irregularities. Both were introduced gradually to avoid abrupt destabilization and to reflect realistic attack behavior that accumulates over time.

Sensor bias drift was implemented by injecting a slowly varying bias into inertial measurement readings. The bias increased progressively during flight while remaining within ranges that do not immediately trigger built-in fault detection mechanisms. This manipulation affects the estimator indirectly and leads to small but persistent discrepancies between predicted and observed motion.

Timing irregularities were introduced by delaying control updates within the simulation loop. Small, variable delays were applied to emulate scheduler jitter or resource contention. These delays did not interrupt control execution but altered the effective timing of the control loop, resulting in gradual degradation of closed-loop behavior.

### D. Flight Scenarios

The evaluation included several representative flight scenarios to capture different operating conditions. These scenarios consisted of stationary hovering, slow translational motion, and simple waypoint tracking. Each scenario was executed under nominal conditions as well as under each degradation attack.

During each run, the stability drift metric and the corresponding early-warning signal were recorded alongside ground-truth state information available from the simulator. This allowed precise assessment of when the warning was triggered relative to observable changes in flight behavior.

## RESULTS

### A. Stability Drift Under Nominal Conditions

Under nominal flight conditions, the stability drift metric remained stable across all evaluated scenarios, including hovering, slow translational motion, and waypoint tracking. Across 20 nominal runs, the aggregated drift metric exhibited a mean value of 0.016 with a standard deviation of 0.006, reflecting sensor noise and minor modeling inaccuracies without any sustained growth.

As shown in Fig 3, the dynamic threshold computed from nominal baseline statistics was never exceeded during normal operation. Importantly, no early-warning events were triggered during aggressive but non-degraded maneuvers, indicating that the proposed metric is robust to benign dynamic variations and does not react to short-term fluctuations. These results establish a stable baseline for the drift metric and confirm that normal flight behavior does not produce false early-warning signals.

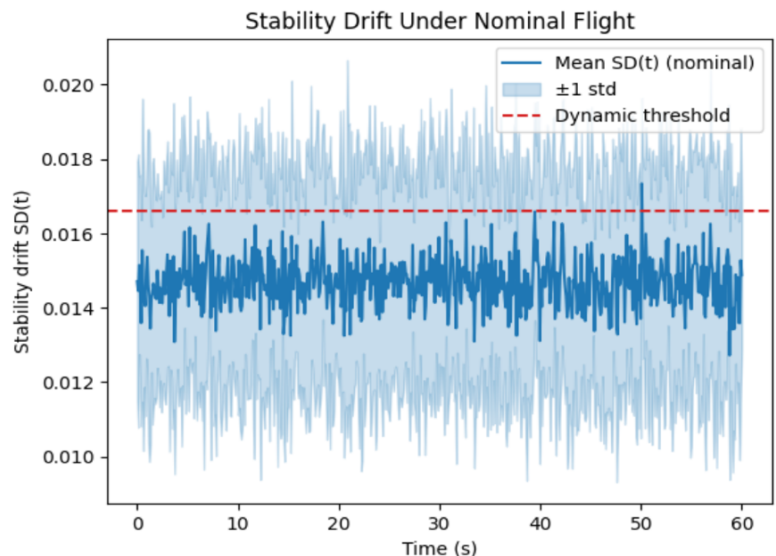

Fig 3. Stability Drift Under Nominal Flight Conditions

### B. Detection of Gradual IMU Bias Drift

When a slowly increasing bias was introduced into inertial sensor measurements, the stability drift metric exhibited a gradual and persistent increase over time. During the early phase of the attack, the drone continued to track its reference trajectory accurately, and estimator residuals remained within normal bounds.

As illustrated in Fig 4, the stability drift metric crossed the detection threshold 6.8s on average before visible instability appeared in the flight trajectory. At the time of detection, the median position error remained below 0.12m, and no oscillatory behavior was observed.

These results demonstrate that stability drift reveals degradation during a pre-instability phase in which conventional trajectory-based or residual-based monitors remain silent.

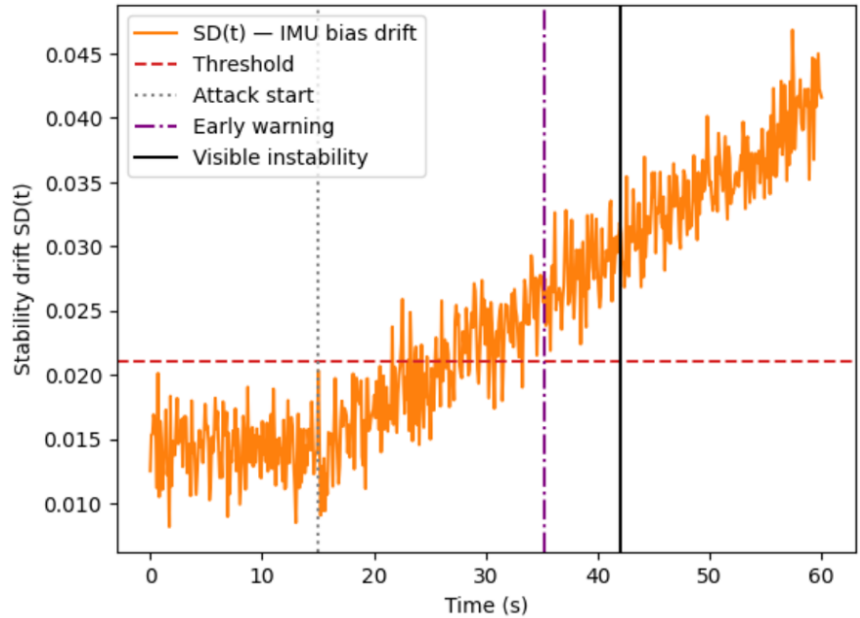

Fig 4. Stability Drift Growth Under IMU Bias Drift

### C. Detection of Timing Irregularities in the Control Loop

Timing irregularities introduced through small, irregular delays in the control loop produced more subtle degradation effects than sensor bias drift. Rather than causing immediate tracking errors, these delays led to delayed control responses and gradual phase shifts in closed-loop behavior.

As shown in Fig 5, the stability drift metric increased consistently and crossed the detection threshold 5.1s on average before the onset of visible oscillations. Detection performance remained consistent across hovering and waypoint-tracking scenarios, demonstrating sensitivity to degradation effects that are difficult to detect using conventional timing or residual-based monitors.

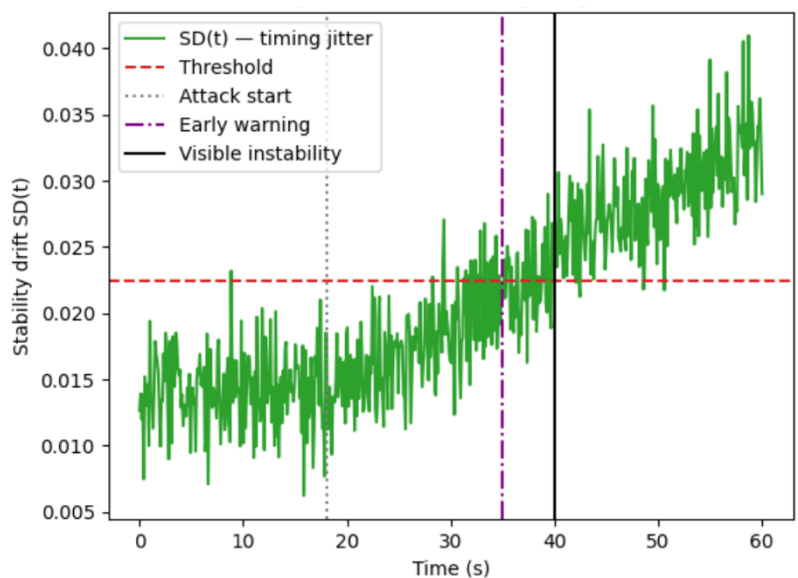

Fig 5. Stability Drift Under Timing Irregularities

**D. Detection Timing Across Flight Scenarios**

To evaluate robustness across operating conditions, detection performance was analyzed under hovering, slow translational motion, and waypoint-tracking scenarios. In all cases, early-warning signals were raised prior to visible instability.

Across all scenarios and attack types, detection lead times ranged from 4.9s to 7.3s, with low variance across repeated runs. This consistency indicates that the proposed method does not rely on a specific flight mode or dynamic regime.

**E. False Positives and Robustness**

No false early-warning events were observed during nominal flights or during aggressive but non-degraded maneuvers. The use of windowed aggregation and dynamic thresholding prevented transient disturbances from triggering alarms.

This robustness is critical for practical deployment, as it enables early detection of gradual degradation without increasing false positives during normal operation.

**CONCLUSION**

This work presented an early-warning approach for detecting gradual degradation in UAV control systems. Instead of reacting to visible instability or abrupt faults, the proposed method monitors the divergence between predicted and observed state transitions and identifies stability drift at an early stage. This allows degradation to be detected while the drone is still flying normally and before oscillations or large tracking errors appear.

The approach was evaluated on a PX4 x500 platform in a software-in-the-loop environment under two realistic degradation scenarios: gradual IMU bias drift and timing irregularities in the control loop. In both cases, the stability drift metric provided a consistent early warning several seconds before visible instability occurred. During this early-warning phase, the drone remained controllable and trajectory errors were still small, highlighting the practical value of detecting degradation before it becomes obvious.

An important aspect of the proposed method is that it operates externally to the flight stack and relies only on standard telemetry. This makes it suitable for deployment without modifying the autopilot firmware or estimator internals. At the same time, the use of windowed aggregation and dynamic thresholding prevents false alarms during nominal flight and aggressive but non-degraded maneuvers. Overall, the results show that stability drift can serve as a reliable indicator of emerging instability caused by slow attacks or faults. While the method does not identify the attack source or directly prevent failure, it provides a useful time margin that can be used to trigger mitigation actions or transition the system into a safer operating mode. Future work will explore integrating the early-warning signal with adaptive control responses and extending the evaluation to hardware flight experiments and additional forms of degradation.

**DICUSSION**

This work focuses on a class of problems that are often overlooked in UAV security and monitoring: gradual degradation that develops without triggering obvious alarms. The results suggest that such degradation affects the underlying dynamics of the system before it becomes visible in trajectories or estimator residuals. By monitoring divergence between predicted and observed behavior, stability drift provides a way to observe this hidden phase of system evolution.

One important implication of this approach is that it shifts the role of monitoring from fault confirmation to early awareness. Traditional mechanisms are designed to react once instability is already present, which limits the range of possible responses. In contrast, early-warning signals can be used to initiate conservative actions while the system is still operating normally. This does not require immediate intervention, but it gives higher-level components or operators more time to make informed decisions. The difference in response between sensor-related degradation and timing-related degradation highlights another strength of the approach. These two attack types affect the system in different ways, yet both produce a consistent increase in stability drift. This suggests that the method captures a general property of system behavior rather than relying on specific signatures. As a result, it can complement existing detectors that are tuned to particular fault or attack models.

The external design of the monitoring system is also significant. Because the method relies only on standard telemetry, it avoids tight coupling with the flight stack and does not depend on access to estimator internals. This makes the approach easier to deploy and reduces the risk of interfering with safety-critical control logic. At the same time, this design choice introduces limitations, as the simplified prediction model cannot fully represent the internal behavior of the autopilot under all conditions.